\documentclass[twocolumn,           
               showpacs,            
               preprintnumbers,     
               aps,                 
               prd,                 
               superscriptaddress,  
               nofootinbib,         
               tightenlines,        
               floats,floatfix,      
               ]{revtex4-1}

\usepackage{graphicx}
\usepackage{dcolumn}
\usepackage{bm}
\usepackage{latexsym}
\usepackage{amsmath,amssymb}        
\usepackage{color}
\usepackage{psfrag}
\usepackage{natbib}

\begin{document}

\title{Unification models with reheating via Primordial Black Holes}

\author{J. C. Hidalgo}
 \email{jhidalgo@astro.unam.mx}
\affiliation{%
  Instituto de Astronom\'{\i}a, UNAM, Ciudad Universitaria, 04510,
  M\'exico D.F., M\'exico.}  
\affiliation{%
  Depto. de F\'{\i}sica, Instituto Nacional de Investigaciones
  Nucleares, Apdo. Postal 18-1027, 11801, M\'exico D.F., M\'exico.}  

 \author{L. Arturo Ure\~na-L\'opez}%
 \email{lurena@fisica.ugto.mx}
 \affiliation{%
   Departamento de F\'isica, DCI, Campus Le\'on, Universidad de Guanajuato,
   37150, Le\'on, Guanajuato, M\'exico.}

\author{Andrew R. Liddle}
 \email{a.liddle@sussex.ac.uk}
 \affiliation{%
   Astronomy Centre, University of Sussex, Brighton BN1 9QH, United Kingdom}
 
\date{\today}

\begin{abstract}
We study the possibility of reheating the universe
through the evaporation of Primordial Black Holes created at the end
of inflation. This is shown to allow for the unification of inflation
with dark matter or dark energy, or both, under the dynamics of a single scalar field. We
determine the necessary conditions to recover the standard Big Bang by
the time of nucleosynthesis after reheating through black holes.
\end{abstract}

\pacs{04.70.-s,98.80Cq,98.80Jk}

\maketitle

\section{Introduction \label{sec:introduction}}
The quest for the unification of inflation, dark matter, and dark
energy, in different combinations, by a single field, has been studied
in
Refs.~\cite{Peebles:1998qn,Lidsey:2001nj,Padmanabhan:2002sh,Scherrer:2004au,Matos:2005yt,Arbey:2006it,Cardenas:2006py,*Cardenas:2007xh,*Panotopoulos:2007ri,Liddle:2006qz,*Liddle:2008bm,Lin:2009ta,Henriques:2009hq,Bose:2009kc,delaMacorra:2009bi,*delaMacorra:2009yi,BasteroGil:2009eb,Lerner:2009xg,*Okada:2010jd,DeSantiago:2011qb}.
The main motivation behind all proposals is that we do not yet
understand the nature of the components responsible for the three
phenomena, but we do know that their special properties are beyond the
realm of the ordinary matter described by the standard model of
particle physics.

An extreme, most economical, possibility is that all three phenomena
can be explained by the existence of one single field. As was first
put forward in Ref.~\cite{Liddle:2006qz,*Liddle:2008bm}, the simplest
option at hand is a scalar field $\phi$ with a potential of the form
$V(\phi) = V_0 + (1/2) m^2 \phi^2$. The energy scale $V_0$ is to be
set at the tiny value of the observed cosmological constant considered
in the concordance $\Lambda$CDM model, and the mass scale of the
field, $m \simeq 10^{-6} m_{\rm pl}$, is determined by the amplitude
of primordial perturbations generated during inflation.

A crucial ingredient in unification schemes that include the inflaton
field is a proper and smooth transition from an inflationary era to
the Hot Big Bang phase we are all familiar with. There needs to be a
\emph{reheating} phase in which the inflaton field can transfer most
of its energy to other relativistic particles, yet be able to survive
the reheating process and attain the right amplitude to later become a
subdominant {dark matter/dark energy} component in the
radiation-dominated era.

The viability of some phenomenological models of reheating for
unification models was explored in
Refs.~\cite{Lidsey:2001nj,Liddle:2006qz,Liddle:2008bm}. Some other,
more detailed proposals have been explored in the literature, from
extra periods of inflation at low energy scales~\cite{Liddle:2008bm},
to the more popular gravitational particle production at the end of
inflation~\cite[e.g.][]{Ford:1977dj,Spokoiny:1993kt,Peebles:1998qn,Huey:2001ae,*Dimopoulos:2001ix,Bose:2009kc}.
This reheating mechanism is largely ineffective, however, because the
relative proportion of radiation to scalar field  $\rho_{\rm rel}/
\rho_{\phi} $ evolves at the same rate as the proportion of
gravitational waves to scalar field $\rho_{\rm
  GW}/\rho_{\phi}$~\cite{Sahni:2001qp}. In the case of steep
potentials, inflation is followed by a kination era (a possibility
first explored in the context of electroweak baryogenesis
\cite{Joyce:1996cp}).
If the kinetic energy of the scalar field dominates for some time,
the relative density of gravitational waves created during inflation
can grow significantly and violate the observational bounds.
Specifically, the process of Big Bang Nucleosynthesis (BBN) imposes a
condition on the energy density of gravitational waves $\rho_{\rm GW}$, 
BBN requiring that~\cite{Shvartsman:1969mm,*Carr:1980ab,*Cyburt:2004yc,*Carr:2009jm}
\begin{equation}
\label{gw:condition}
\rho_{\rm  GW}(t_{\rm BBN}) < 0.2 \, \rho_{\rm rel}(t_{\rm BBN}) \, .
\end{equation}
In fact, the BBN bound in Eq.~(\ref{gw:condition}) has been used to
constrain and exclude some models of braneworld inflation and other
unification models~\cite{Giovannini:1998bp,*Boyle:2007zx}.

In this paper, we explore an alternative mechanism to reheat the
Universe, exploiting the evaporation of primordial black holes (PBHs)
produced at the end of the inflationary
phase~\cite{Barrow:1990he,GarciaBellido:1996qt}, and study its 
usefulness in unification models. A small initial population of tiny
black holes will grow significantly (in relative energy density) in a
universe dominated by a stiff fluid. The evaporation of such black
holes at the right time would produce the required amount of radiation
more effectively than, for instance, the process of quantum particle
production in an expanding space-time.

In the PBH reheating scenario, the black hole component effectively
reduces the relative density of gravitational waves as   
\begin{equation}
\label{rhogw:rhopbhs}
\frac{\rho_{\rm GW}}{\rho_{\rm PBH}} \propto \left(\frac{a}{a_{\rm
    end}}\right)^{-1}. 
\end{equation}
Reheating takes place when the energy density $\rho_{\rm PBH}$ is
transformed into relativistic particles. Therefore, even if the
densities $\rho_{\rm GW}$ and $\rho_{\rm PBH} $ were equivalent at the
end of inflation, we would only require 3 $e$-folds of kinetic energy
domination to meet the nucleosynthesis requirements. PBH reheating
therefore solves the problem described above. Another advantage of
reheating through the evaporation of PBHs is that the unification
field does not need to be coupled to other matter fields, hence it
can later behave as a totally uncoupled dark matter component. PBH
reheating for unification purposes was first considered in
Ref.~\cite{Lidsey:2001nj}; see also Ref.~\cite{BasteroGil:2009eb}. 

Our aim here is to determine the general conditions under which
{the unification of inflation with either, or both, of the dark
  fields} can achieve the appropriate reheating with the aid of PBHs
and set up the required conditions before BBN.

The featured model of PBH reheating is described in detail in
Sec.~\ref{sec:pbh-reheating}. Together with this description, we
establish the conditions that the scalar field potential must fulfil
in order to undergo a successful PBH reheating. A particular example
of our proposal is illustrated in Sec.~\ref{sec:examples}. Finally, in
Sec.~\ref{sec:remarks}, we frame our proposal in the context of
unification models and draw some concluding remarks.

\section{PBH Reheating} 
 \label{sec:pbh-reheating}

There are three basic requirements for a successful PBH reheating: 
\begin{itemize}
  \item[i)] a significant number of PBHs produced at the end of
    inflation,
  \item[ii)] that PBHs become the dominant component in the Universe,
    and
  \item[iii)] the evaporation of the PBHs after they become the
    dominant component and before the time of BBN. 
\end{itemize}
The sequence described above introduces some tuning. We require the
formation of black holes massive enough to last for a few e-folds
while the Universe is dominated by the scalar field kinetic
energy. Additionally, the same PBHs must evaporate before BBN, or
before any prior well-tested process at higher energies in the
standard Big Bang model. Each one of these conditions can be used to
constrain the free parameters of the reheating process.

\subsection{Production of PBH and domination}
In recent years, several works have demonstrated the compatibility of
the observed power spectrum of matter fluctuations at cosmological
scales with the formation of a significant number of PBHs at the end
of inflation ($t_{\rm end}$), see
e.g.~Ref.~\cite{Chongchitnan:2006wx,*Kohri:2007qn,*Alabidi:2009bk,*Zaballa:2009xb}, or
indeed due to preheating~\cite{Green:2000he}. Moreover, the relative
growth of $\rho_{\rm PBH}$ with respect to other components is largest
for PBHs formed at $t_{\rm end}$. This justifies considering the black
holes formed right after the end of inflation as those responsible for
reheating. {PBHs could also be originated by features in the
  inflationary potential. In the inflationary phase, these features
  break the slow-roll conditions momentarily, and may enhance the amplitude
  of fluctuations and black hole production at a specific scale after
  the end of inflation~\cite[e.g.][]{Ivanov:1997ia}. This presents an
  alternative to the production mechanism considered here.}

To quantify the number of black holes at the end of inflation, it is
customary to define with $\beta$ the initial mass fraction of PBHs at
the time of formation $t_{\rm end}$. The Press--Schechter formalism of
structure formation prescribes that $\beta$ is a function of the
variance $\sigma$ of matter fluctuations. If we define $\mathbb{P}$ as
the Gaussian probability distribution of $\delta$, with $\sigma$ its
variance at that particular scale, then
\begin{equation}
  \beta(\sigma) = f_M \int^\infty_{\delta_{\rm min}}
  \mathbb{P}(\delta,\sigma) d\delta \, ,
  \label{def:beta}
\end{equation}
where $\delta_{\rm min} \simeq w_{\rm
  end}$~\cite{carr:1975,*Musco:2004ak}, and $w_{\rm end} =
w_{\phi}(t_{\rm end})$ 
is the equation of state at the end of inflation, which coincides with
that of the inflaton field which is by then the dominant matter
component. Here, $f_{\rm M} \approx 3^{-3/2}$ is the fraction of the
horizon mass which constitutes the black hole at the time of
formation~\cite{Sanz:1985np}. The amplitude of threshold
inhomogeneities $\delta_{\rm min}$ is determined by the equation of
state of the dominant energy component at the end of inflation, $w_{\rm
  end}$.\footnote{Note that we evaluate $w_{\rm end}$ at times
  shortly after the end of inflation, where $w$ can
  have a value between $w_{\rm inf} = -1/3$ right at the end of inflation,
  and the maximum at the kinetic energy domination $w = 1$. Also, the
  precise value of $\delta_{\rm min}$ is dependent of the shape of the
  initial configuration~\cite{Green:2004wb,*Hidalgo:2008mv}.}

The energy density of the produced PBHs evolves as pressureless matter
(i.e.\ dust, with $w_{\rm PBH} = 0$), so that after time $t_{\rm end}$, the
ratio of the PBHs to the scalar field energy densities is
\begin{subequations}
\label{omega:beta}
\begin{eqnarray}
  \frac{\rho_{\rm PBH}}{\rho_{\phi}} &=& \frac{\beta(\sigma)}{1 -
    \beta(\sigma)} \left( \frac{a(t)}{a_{\rm end}}
  \right)^{\frac{\Delta(N)}{N}} = \frac{\beta(\sigma)}{1 -
    \beta(\sigma)} e^{\Delta(N)} \, , \;\; \\
  \Delta(N) &=&  {3} \int^N_0 w_{\phi}(N^\prime) dN^\prime \, ,
\end{eqnarray}
\end{subequations}
where $N$ is the number of $e$-folds elapsed from $t_{\rm end}$,
$\sigma$ is the mean amplitude of matter fluctuations, and $\Delta(N)$
is the (exponential) growth factor of the PBHs.
Eq.~(\ref{omega:beta}) takes into account the fact that typically the
scalar field equation of state does not remain constant after the end
of inflation.

The mass fraction of PBHs will always increase as long as $\Delta(N) >
0$, that is:
\begin{equation}
  \qquad w_{\phi} > 0, \qquad\mathrm{for}\,\, \mathrm{some}\quad 
  t_{\rm end}\, <\, t\,<\, t_{\rm kin} \,,
\label{eos:condition}
\end{equation} 

\noindent {where $t_{\rm kin}$ is the time when the kination era of the scalar
field ends and it starts to behave like dark matter or dark
energy. For definiteness, we take the equation of state at the
end of the kination era to be $w_{\rm kin} := w_{\phi}(t_{\rm kin})~=~
0$, and the total growth factor is $\Delta(N_{\rm kin})$. So far,
we are assuming that, as in most cases, the dark field-like behaviour
of the inflationary field appears at the onset of oscillations of the
field around the minimum of the scalar potential. This need not
necessarily be the case, but the final results are insensitive to the
precise details of the dark matter/dark energy transition.}

It is also clear, from Eq.~(\ref{omega:beta}), that for PBH domination
we require
\begin{equation}
  \label{eq:domination}
  \ln \beta(\sigma) > - \Delta(N_{\rm kin})  \, .
\end{equation}
Whether this condition is satisfied will depend upon the scalar field
model and the exact evolution of its equation of state in the
post-inflationary era. One can foresee, though, that some models may
fail to satisfy condition~(\ref{eq:domination}).

Finally, we note that in the PBH reheating process we do
not expect a significant PBH merging rate at the phase of
domination. This is because the value of $\beta$ is initially much
smaller than $1$, and the density of PBHs grows predominantly because of the
rapidly diluting scalar field energy density~\cite{Anantua:2008am}.

\subsection{The inflationary energy scale}
The dominant radiation at the start of the Big Bang is generated by
the evaporation products of PBHs. Depending on the initial mass of the
black holes, the time of evaporation, $t_{\rm ev}$, can be shorter than,
equal to, or larger than the time at which $\phi$ {starts behaving as
dark matter/dark energy}, $t_{\rm kin}$ (the relation between these times
can be derived directly from the characteristics of the scalar field
model under consideration). 

For the sake of clarity, let us consider the case in which
$t_{\rm ev} = t_{\rm kin}$.
The
evaporation time of PBHs is determined by their mass, which is directly
related to the energy scale at the end of inflation, $\rho_{\rm end}
\simeq V_{\rm end}$,
\begin{equation}
  M_{\rm PBH} = f_M \, \frac{4}{3}\pi \rho_{\rm end} H^{-3}_{\rm
      end} =   4 \pi \sqrt{3} f_M\left(\frac{\rho_{\rm
        end}}{m_{\rm pl}^4}\right)^{-1/2} m_{\rm pl} \, ,
\label{pbhmass:density}
\end{equation}
where $m_{\rm pl}$ is the Planck mass. The time of evaporation is
given by~\cite{Hawking:1974sw,Kim:1999iv} 
\begin{equation}
\label{eq:tev}
t_{\rm ev } = 6.3 \times 10^{-41} \Phi^{-1}(M_{\rm PBH}) \left(\frac{M_{\rm
      PBH}}{m_{\rm pl}} \right)^{3} \mathrm{sec.}
\end{equation}
Here $\Phi(M_{\rm PBH})$ is a function of the directly-emitted
species of particles and takes a value of order $10$ for the PBH
masses of interest~\cite[e.g.][]{Kim:1999iv}. Combining Eqs.~(\ref{pbhmass:density}) 
and~(\ref{eq:tev}), we find that
\begin{equation}
  t_{\rm ev} = 
  2.3 \times 10^{-41}\Phi^{-1}(M_{\rm PBH}) \left( \frac{\rho_{\rm
        end}^{1/4}}{m_{\rm pl}} \right)^{-6} \mathrm{sec} \,
  . \label{rhoend:time}
\end{equation}
This time must also be the beginning of the standard Hot Big Bang
(HBB), at some point before BBN. This last requirement imposes a
bound on the initial mass of the black holes and, consequently, on
$\rho_{\rm end}$. Indeed, in view of Eq.~(\ref{rhoend:time}), the
condition $t_{\rm ev} \leq t_{\rm BBN} = 1\, \mathrm{sec.}$ implies
\begin{equation}
  \label{onesec:evap}
  V^{1/4}_{\rm end} \geq 1.6 \times 10^{-7} \Phi^{1/6}(M_{\rm PBH}) m_{\rm pl}
  \, .
\end{equation} 
\subsection{The right amount of radiation}
On the other hand, we can determine whether radiation domination is
achieved after the evaporation of PBHs. Assuming again that the
standard HBB is recovered at the time of evaporation, we can write a
simple expression for the ratio of PBH to scalar field densities as
\begin{eqnarray}
  \left(\frac{\rho_{\rm PBH}}{\rho_{\phi}}\right)_{\rm ev} &=& \left(
    \frac{\rho_{\rm rel}}{\rho_{\rm dm}}\right)_{\rm ev} \simeq
  \left(\frac{g_{\rm ev} t_{\rm eq}}{g_{\rm eq} t_{\rm
        ev}}\right)^{1/2} \nonumber \\ 
  & \simeq &
  3\times 10^{26} \left(\frac{V^{1/4}_{\rm end}}{m_{\rm
        pl}}\right)^{3} \, , \label{densities:tosc}
\end{eqnarray}
where we have considered that from $t_{\rm ev}$ onwards the
Universe is radiation dominated, and $g_{\rm ev}$ ($g_{\rm eq}$) are
the relativistic energy degrees of freedom at that time of evaporation
(matter--radiation equality), respectively.

Radiation domination is ensured as long as the evolution of the scalar
field equation of state allows the matching between
Eqs.~(\ref{omega:beta}) and~(\ref{densities:tosc}) at the time of PBH
evaporation. We will see that this is possible even when the
constraints on the products of PBH evaporation are considered. 

\subsection{Additional constraints}
There are two important bounds on the density of the products of
evaporation of PBHs. These can be written in terms of the energy
density at the time of black hole formation and, therefore, introduce
bounds to the unification inflationary potential. The first constraint
comes from the consideration that a PBH could leave behind a Planck
mass relic after evaporation. Such relics must not exceed in density the dark
matter component~\cite{Josan:2009qn,*Carr:2009jm}, i.e., we require   
\begin{equation}
  \label{relic:cond}
  \frac{\rho_{\rm relic}}{\rho_{\rm DM}} < 1\,.
\end{equation}
From Eq.~(\ref{pbhmass:density}) we can read the proportion between
the PBH mass and the Planck mass of its relic. Therefore the energy
density ratio of PBHs with relics is
\begin{equation}
  \label{rhorad:rhorelic}
 \left(\frac{\rho_{\rm PBH}}{\rho_{\rm relic}}\right)_{{\rm ev}} =
 \left(\frac{\rho_{\rm rel}}{\rho_{\rm relic}}\right)_{{\rm ev}}  = 
  4 \pi \sqrt{3} f_M \left(\frac{m_{\rm pl}}{V^{1/4}_{\rm end}}\right)^2 \, .
\end{equation}
Combining Eqs.~(\ref{densities:tosc}) and~(\ref{rhorad:rhorelic}), the
condition~(\ref{relic:cond}) is met when
\begin{equation}
\label{relic:vcond}
V^{1/4}_{\rm end} \lesssim 6.8\times 10^{-6} m_{\rm pl}\, .
\end{equation}
If the condition~(\ref{relic:cond}) is met after
evaporation, then the same ratio of relics to dark matter is maintained
at subsequent times. 

The combination of Eqs.~(\ref{onesec:evap}) and~(\ref{relic:vcond})
leaves but a narrow window for the range of energies at which
inflation must end if we want the PBHs to reheat the Universe. The
tight constraint can however be relaxed if no Planck mass relics are
left over after evaporation~\cite{Hsu:2006pe,*Carr:2011pr}.

This is a crucial point for standard (slow-roll) inflationary
scenarios, for which the end of inflation takes place at energies
$V_{\rm end}^{1/4} \sim 10^{-3} m_{\rm pl}$.  PBH reheating in this
case is only possible if no Planck relics are left over after
evaporation, or if the black holes responsible for reheating are
created at energies below $\rho_{\rm end}^{1/4}$.

A second constraint comes into play in the form of gravitational
waves. When black holes evaporate, a considerable fraction of energy
is emitted in the form of gravitons. Paradoxically for the initial
motivation of this paper, it appears that in the case of a dominant
energy density component of PBHs, a significant amount of
gravitational waves is produced at evaporation. This
could violate the very constraint we intend to alleviate.

When PBHs dominate and then evaporate, the amount of gravitational
radiation produced is~\cite{Anantua:2008am}, 
\begin{equation}
  \Omega_{\rm GW}  =  0.36 / g_{\rm ev}.
\end{equation}
We are interested, however, in times of evaporation prior to
nucleosynthesis, when $g_{\rm ev}\geq 10.75$. Consequently,
\begin{equation}
\label{eq:pbh-gw}
  \left(\frac{\rho_{\rm GW}}{\rho_{\rm rel}}\right)_{{\rm ev}} \sim 
  \Omega_{\rm  GW} (t_{\rm ev})\lesssim 0.03 \, ,
\end{equation}
which readily satisfies the gravitational waves bound in
Eq.~(\ref{gw:condition}). 

\section{Examples}
\label{sec:examples}

To illustrate our proposal, let us look at some scenarios where PBH
reheating can take place. In the model considered in
Ref.~\cite{Lidsey:2001nj}, {which intended a unification into
  a single scalar field of inflation and dark matter only}, as in many
other braneworld models, inflation ends at an energy scale of
$V^{1/4}_{\rm end} \approx 10^{-6} m_{\rm pl}$, which is within the
allowed values of constraints~(\ref{onesec:evap})
and~(\ref{relic:vcond}). The expansion of the Universe between the end
of inflation and the onset of oscillations is such that
\begin{equation}
  \frac{a_{\rm kin}}{a_{\rm end}} = 60 \approx e^4 \, ,
\end{equation}
and the scalar field enters a kination phase in between, during which
$w_\phi \simeq 1$. This means that the growth factor in
Eq.~(\ref{omega:beta}) is $\Delta(N_{\rm kin}) \simeq 12$. In
consequence, Eq.~(\ref{eq:domination}) tells us that to reach the
domination of PBHs before $\phi$ starts oscillating, we must have
$\beta > 10^{-6}$. This is in agreement with the results in
Ref.~\cite{Lidsey:2001nj}.

{Another example is the quintessential inflation model
in~\cite{Peebles:1998qn}, for the unification of inflation and dark
energy. After inflation, the field goes into a kination era for which
the Universe expands $a_{\rm kin}/a_{\rm end} = 10^{8} \approx
e^{18}$. 
Thus, the quintessential model can be more easily
implemented, as it only requires $\beta > 10^{-24}$ to generate enough
radiation from PBHs. 
Moreover, if the inflation ends due to an instability of the
hybrid inflation kind, then the smaller-scale fluctuations would show an
enhanced amplitude~\cite{Roberts:1994ap}, rendering a larger probability
of PBH formation.}

\begin{figure}[!t]
    \psfrag{w-=0.6-w-eoi-=0.02}{\scriptsize${w=.6, w_{\rm end}=.02}$}
    \psfrag{w-=1.0-w-eoi-=0.02}{\scriptsize${w= 1, w_{\rm end}=.02}$}
    \psfrag{w-=1.0-w-eoi-=0.10}{\scriptsize${w= 1, w_{\rm end}=0.1}$}    
    \psfrag{w-=0.6-w-eoi-=0.10}{\scriptsize${w=.6, w_{\rm end}=0.1}$}
    \psfrag{E}{\small$\rho_{\rm ev}^{1/4}$}
    \psfrag{F}{\small$\rho_{\rm BBN}^{1/4}$}
    \includegraphics[width=0.49\textwidth]{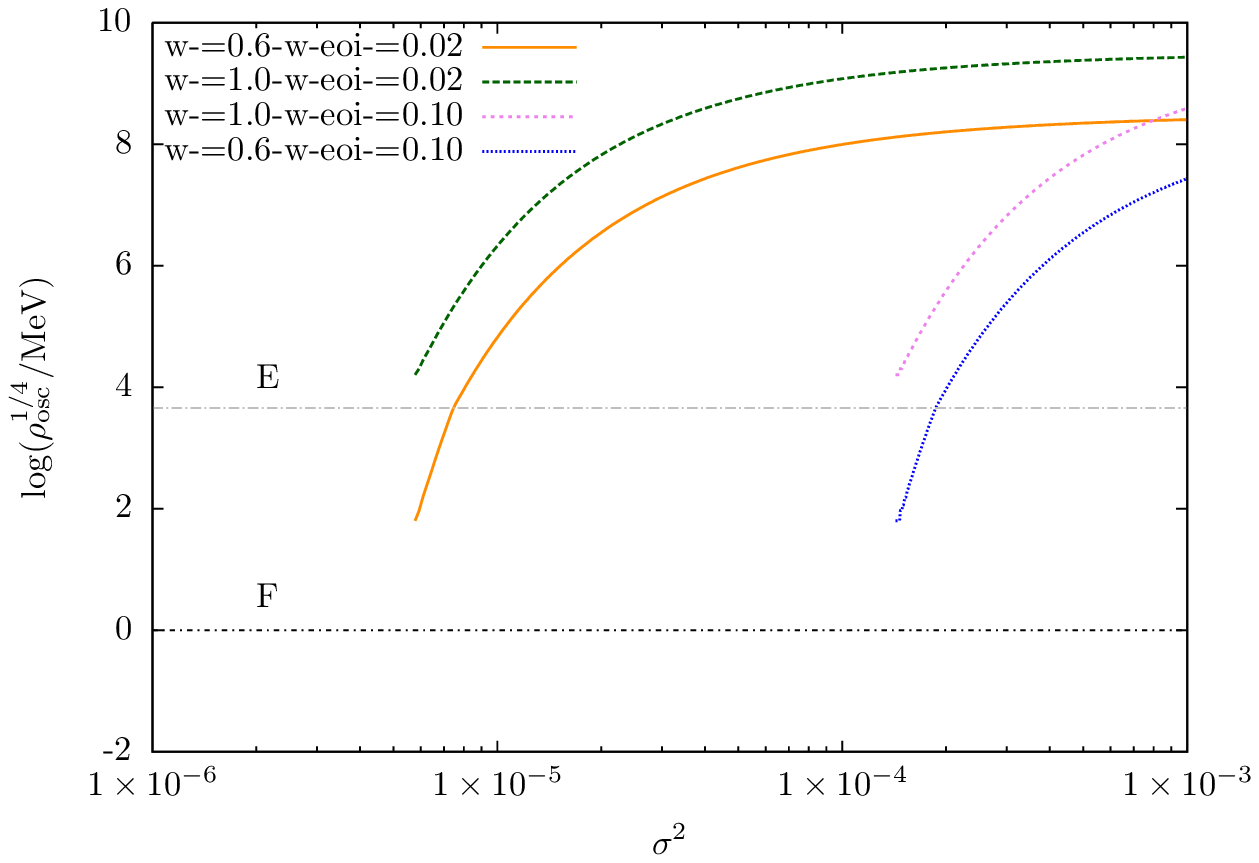}
    \includegraphics[width=0.49\textwidth]{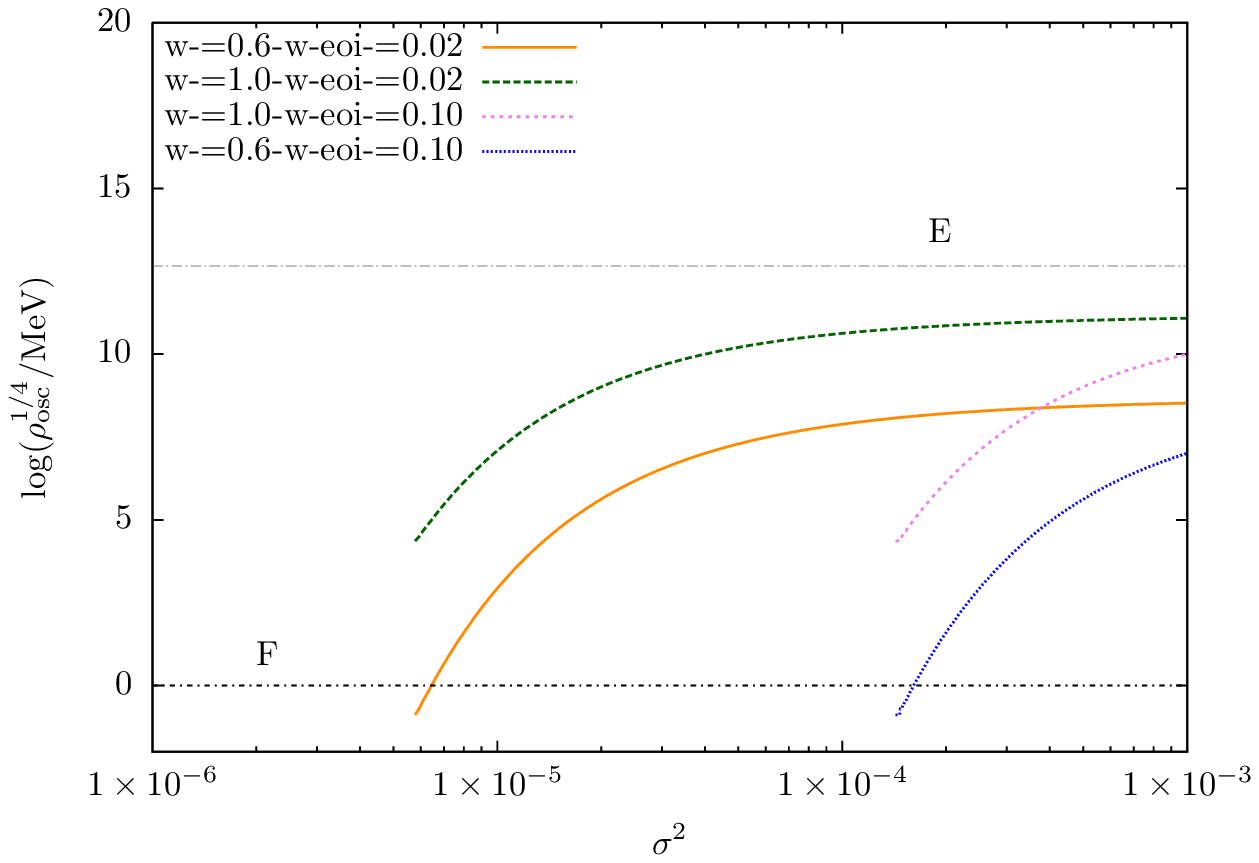}
    \caption{The curves indicate the required variance of the matter
      spectrum $\sigma^2$, see Eq.~(\ref{def:beta}), to create enough
      PBHs that subsequently evaporate and account for all the
      radiation in the early Universe before BBN. The figure at the
      top considers $\rho_{\rm end} = (10^{-6} m_{\rm pl})^4$ and the
      bottom figure takes $\rho_{\rm end} = (10^{-3} m_{\rm
        pl})^4$. We plot the scales of the evaporation of
      PBHs and of BBN with a grey and a black line, respectively. The
      area below $\rho_{\rm BBN}$ is excluded by
      observations.}
    \label{fig:1}
\end{figure}


In Fig.~\ref{fig:1} we present graphically the valid parameter values
for a successful PBH reheating. The curves in both figures indicate
the mean amplitude of matter fluctuations, $\sigma$ in
Eq.~(\ref{def:beta}), required for a given energy density (or time) at
which the scalar field starts oscillating.

The figures show the importance of the difference between the equation
of state right after the end of inflation $w_{\rm end}$ and at the
subsequent kination epoch $w_{\phi}$. For simplicity, we have assumed
that $w_\phi = {\rm const.}$ during the kination phase of the scalar
field. This is not a strong restriction, as the total growth factor
can be written in terms of a mean equation of state as $\Delta(N_{\rm
  kin}) = 3 w_{\rm mean} N_{\rm kin}$, and then the plots in
Fig.~\ref{fig:1} can be interpreted in terms of $w_{\rm mean}$ too.

In particular, we see that the small value $w_{\rm end} \leq 0.1$
allows the formation of large populations of PBHs with a relatively
small variance. Note that the potential of the figure at the bottom
violates the constraint imposed by Planck relics, a constraint that,
as mentioned above, need not necessarily apply.

By considering a range of values for $\rho_{\rm kin}$ above (below) the
evaporation scale $\rho_{\rm ev}$ in the top (bottom) figure, we are
showing that the conditions required for PBH reheating when $t_{\rm
  ev} \neq t_{\rm kin}$ are not too different from those in the case of
equality (described in the previous section).

For the sake of generality we are not computing here the amplitude of
fluctuations for specific models at the end of inflation. It suffices
to mention that for some models with $w_{\rm end} \approx 0$,
  and under suitable conditions, the production of a considerable
  amount of PBHs is possible at horizon and subhorizon 
  scales~\cite{Lyth:2005ze,*Suyama:2006sr,*Jedamzik:2010dq,Lyth:2010zq}.


\section{Unification models and final remarks} \label{sec:remarks}

If the unification of inflation with dark matter or dark energy, {in
  any given combination}, is to be achieved by a single field, then an
efficient reheating process is needed which does not heavily rely upon
the decay of the unification field.

We have seen that, in the presence of a large enough population of
PBHs, unification models can be possible if the unification field
enters a long enough {kinetic-dominated} era, so that the PBHs
come to dominate the matter budget. PBH domination is easier for
{long kination periods} and stiffer values of the effective
equation of state $w_{\phi}$, but the latter also requires a larger
value of the mean amplitude of primordial perturbations $\sigma$ at
the end of inflation.

In this article we have not included an explicit mechanism for the PBH
formation, as would be required for a fully self-contained
model. Most economical, if possible, would be for the unification
field itself to feature perturbations rapidly growing at late times,
as may happen if its second derivative becomes large as inflation
ends. Alternatively, a phase transition (for instance of hybrid
inflation form) may lead to enhanced perturbations as it takes
place~\cite[e.g.][]{Roberts:1994ap,GarciaBellido:1996qt,Bugaev:2011qt,*Bugaev:2011wy,*Lyth:2012yp}.
{In any case, it is worth mentioning that the mechanism required for
  PBH formation need not spoil or interfere with the dynamics followed
  by a unification model in the dark matter or dark energy eras.}

Because PBHs form right after the end of inflation, most of the
observational restrictions can be written in terms of the energy scale
at the end of inflation. In the simplest PBH scenario, the constraints
are satisfied only if this energy scale is a few orders of magnitude
below the Planck value. One should bear in mind that the possibility
of production of Planck relics renders PBH reheating incompatible with
standard inflationary models, unless the energy scale at the end of
inflation happens to be of the order of $10^{-6} \, m_{\rm Pl}$.

The strongest constraint on the featured mechanism comes from the
evolution of the scalar field equation of state after inflation. The
unification model should have a long enough kination period to allow
PBH domination before evaporation. Typical examples are braneworld
inflation with an exponential
potential~\cite{Copeland:2000hn,Lidsey:2001nj} 
{and quintessential inflation~\cite{Peebles:1998qn}}. However,
even if the Planck relic constraint could be set aside, this is a
condition that many inflationary models will not be able to surpass.

Our results rely heavily on the assumption that reheating proceeds
from PBHs formed right at the end of inflation. The results can change
if PBHs are formed after the end of inflation, or if we consider the
kination period to happen beyond BBN (a possibility explored in
Ref.~\cite{Dutta:2010cu}). Therefore, the parameters of the theory can
be flexible beyond the constraints presented here.

\begin{acknowledgments}
J.C.H.\ gratefully acknowledges sponsorship from DGAPA-UNAM and the
support of PAPIIT-UNAM (grant IN116210-3). 
A.R.L.\ was supported by the Science and Technology Facilities Council
[grant number   ST/I000976/1].  
L.A.U.-L.\ thanks the Berkeley Center for Cosmological Physics (BCCP)
for its kind hospitality, and the joint support of the Academia
Mexicana de Ciencias and the United States-Mexico Foundation for
Science for a summer research stay at BCCP. This work was partially
supported by PROMEP, DAIP-UG, and by CONACyT M\'exico under grants
56946, 167335 and I0101/131/07 C-234/07 of the Instituto Avanzado de
Cosmolog\'{\i}a (IAC) collaboration. 
\end{acknowledgments}
 
 
%

\end{document}